# Eight-fold signal amplification of a superconducting nanowire single-photon detector using a multiple-avalanche architecture

Qingyuan Zhao,[1,2] Adam McCaughan,[2] Andrew Dane,[2] Faraz Najafi,[2] Francesco Bellei,[2] Domenico De Fazio,[2,3] Kristen Sunter,[2] Yachin Ivry,[2] Karl K. Berggren[2,*]

[1]*Research Institute of Superconductor Electronics (RISE), School of Electronic Science and Engineering, Nanjing University, Nanjing 210093, China*
[2]*Department of Electrical Engineering and Computer Science, Massachusetts Institute of Technology, 77 Massachusetts Avenue, Cambridge, Massachusetts 02139, USA*
[3]*Department of Electronics and Telecommunications (DET), Polytechnic University of Turin, Corso Duca Degli Abruzzi 24, Torino 10129, Italy*
[*]*berggren@mit.edu*

**Abstract:** Superconducting nanowire avalanche single-photon detectors (SNAPs) with $n$ parallel nanowires are advantageous over single-nanowire detectors because their output signal amplitude scales linearly with $n$. However, the SNAP architecture has not been viably demonstrated for $n > 4$. To increase $n$ for larger signal amplification, we designed a multi-stage, successive-avalanche architecture which used nanowires, connected via choke inductors in a binary-tree layout. We demonstrated an avalanche detector with $n = 8$ parallel nanowires and achieved eight-fold signal amplification, with a timing jitter of 54 ps.

## 1. Introduction

Sueprconducting nanowire single-photon detectors (SNSPDs) are of great interest in fields such as quantum information processing, linear-optics quantum computation [1], and quantum communication [2] due to their high efficiencies, low dark counts, fast recovery times, and low timing jitters [3]. However, one of the limiting factors in their application is the small amplitude of their output pulses, which are on the order of a millivolt. The presence of noise in the readout system can overwhelm these small pulses, and so amplification of the pulse is necessary for accurate readout. For a standard SNSPD architecture of a single nanowire, additional timing jitter [4,5] and dark counts typically result from electric noise if the signal-to-noise ratio (SNR) of its output pulses is low. While external amplifiers can be used to achieve amplification, they ultimately add noise and reduce the SNR of the readout. Spatially multiplexing superconducting nanowire single-photon detectors can realize a single-photon array or a photon-number resolving (PNR) detector [6–8]. However, a scalable readout circuit has not yet realized. Recently, analogous readout architectures which integrated passive electronics with multiple SNSPDs were reported [8,9], using a basic idea of encoding information of photon position or photon number on pulse amplitude. But, scaling up the array or PNR detector was limited by the SNR of a single element which was a typical SNSPD.

To increase the amplitude of the detector output pulses is a direct way to increase the SNR, therefore, superconducting nanowire avalanche photodetectors ($n$-SNAPs, also named cascade-switching superconducting single-photon detectors) [10,11] were developed. The $n$-SNAP is a variation of the SNSPD, with $n$ individual detection nanowires in parallel instead of a single, longer, detection nanowire. The parallelized SNAP architecture increases the total amount of current that the detector carries, and as a result increases the amplitude of the output pulses. A typical $n$-SNAP implementation is shown in Fig.1(a). In an $n$-SNAP, a single photon absorbed in one detection nanowire creates a resistive hotspot, expelling current from that nanowire and redistributing it into to the $n$-1 neighboring detection nanowires. If the $n$-SNAP is biased with enough current, the redistributed current will cause an avalanche in the neighboring nanowires. During this avalanche, the current added to each neighboring nanowire causes it to exceed its critical current ($I_{CN}$), initiating an avalanche of resistive hotspots. Once all the detection nanowires are resistive, the current from all $n$ nanowires is diverted into the output, and so an $n$-SNAP can generate output pulses $n$ times larger than pulses of an SNSPD. Additionally, since this process occurs at the device level, the amplification mechanism does not add external noise and so the SNR scales proportionally with $n$ as well. Unfortunately, the $n$-SNAP design has not scaled well for $n > 4$ detection nanowires [11]. As $n$ is increased, the portion of current that each of the $n$-1 neighboring nanowires receives from the redistributed current shrinks. For an $n$-SNAP biased at $I_B$ with a critical current of $I_C$, the bias condition necessary to guarantee an avalanche of all the detection nanowires into a resistive state is $I_B \geq I_C(N-1)/N$. The minimal $I_B$ that initiates an avalanche is defined as the avalanche current ($I_{AV}$). As $n$ increases, $I_{AV}$ rapidly approaches $I_C$, which can cause problems such as afterpulsing, latching, and increased dark count rates [12].

## 2. Design and fabrication

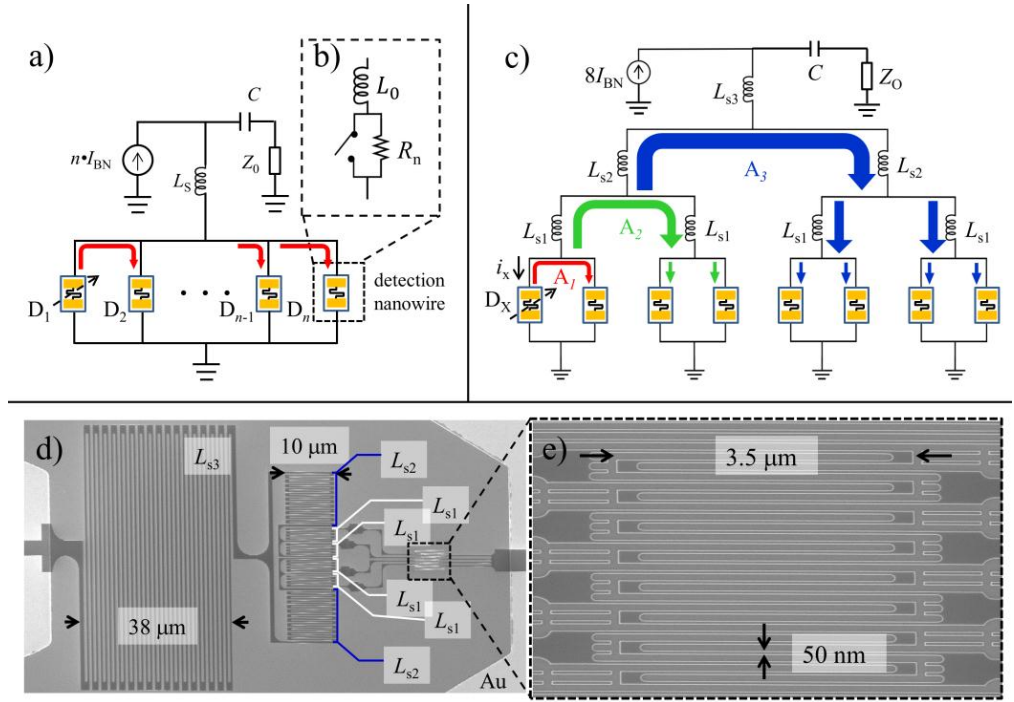

Fig. 1. (a) Schematic diagram of an $n$-SNAP in conventional avalanche architecture. Nanowires from $D_1$ to $D_n$ are paralleled directly. A choke inductor ($L_S$) is connected in series to block the current to leak to the load. (b) Equivalent circuit of a detection nanowire. $L_0$ is the kinetic inductance of the nanowire. $R_n$ is the time-varying resistance as the switch is open by a photon detection. (c) Schematic diagram of an 8-*SNAP. $D_x$ and $i_x$ represent the $x^{th}$ individual nanowire and the current passing through it, respectively. $D_1$ (leftmost detection nanowire) is triggered as the initial condition. The colored arrows represent the current diverted by the sequential avalanches $A_1$ (red), $A_2$ (green) and $A_3$ (blue). (d) SEM images of an 8-*SNAP. The whole device is labeled according to the corresponding circuit parts shown in (b). (e) A close-up of the photon-sensitive individual nanowires, consisting of eight 50 nm wide nanowires.

Here, we introduce a modified $n$-SNAP design, which we refer to as $n$-*SNAP. The $n$-*SNAP uses multiple avalanche stages to increase the total number of nanowires beyond four while maintaining a low avalanche current. Using inductors arranged in a binary-tree layout, we can control the evolution of the avalanche such that the redistributed current from the initial triggered nanowire only needs to overbias one neighboring nanowire at first, but will still cause a resistive avalanche across the whole device. The idea of using multiple parallel SNSPDs was also reported in Ref. [13] as a way to increase the active area of a detector. However, in our design, the configuration of the nanowires and the inductor network is a binary-tree format used specifically to generate a controlled avalanche and increase $n$.

The schematic diagram of the 8-*SNAP is shown in Fig. 1(c) and the SEM pictures of the fabricated device are shown in Fig. 1(d) and (e). In the 8-*SNAP, the eight detection nanowires were arranged to form four pairs, such as $D_1$ and $D_2$. Each pair was connected through a binary tree of choke inductors to form secondary and tertiary avalanches. The arrows in Fig. 1(a) illustrate the order in which the avalanches occur when the detection nanowire $D_1$ is triggered. Current is diverted from $D_1$ and flows into the adjacent nanowire $D_2$, because the choke inductance $L_{S1}$ acts as a high-impedance barrier to the other possible current paths at these short timescales. When $i_2$, the current through $D_2$, exceeds $I_{CN}$, $D_2$ is triggered to resistive state. At this point, both $D_1$ and $D_2$ are resistive, and because we have designed $L_{S2}$ to be larger than $L_{S1}$, most of the current diverted from the $D_1/D_2$ 2-SNAP is routed to the $D_3/D_4$ 2-SNAP through the lowest-impedance path $L_{S1}$. As long as $i_3$ and $i_4$ exceed $I_{CN}$, the second avalanche $A_2$ follows. A similar process then cascades to the four remaining detection nanowires, creating the third avalanche $A_3$. When all the nanowires have switched to the resistive state, the bias current is forced into the external load $Z_0$, generating a voltage pulse for detection.

We simulated the circuit using a electrothermal model and calculated the evolution of the current in each nanowire [14]. After $D_1$ was triggered, the other nanowires were observed to go to resistive in the expected order. The current distribution in each nanowire, as shown in Fig. 2(a), depicts the sequence of the avalanches. Although the current in each nanowire varied during the avalanches, due to the electrothermal feedback in the nanowire [14,15], all of the currents eventually converged and evolved identically after the final avalanche $A_3$ (~40 ps after $D_1$ was triggered).

We designed, fabricated and characterized 8-*SNAPs and also fabricated 2-SNAPs, 3-SNAPs and SNSPDs for comparison. All detectors were made from ultrathin niobium nitride (NbN) film on a single MgO substrate. The critical temperature of the film was 11.4 K and the sheet resistance was 429 $\Omega$/square. Detectors were measured in a probe station at a sample temperature of 2.4 K. The critical current, width, pitch and active area of these devices are: 8-*SNAP (65.0 µA, 50nm, 150nm, $3.5 \times 3.5$ µm$^2$), 2-SNAP (15.5 µA, 50nm, 150nm, $3.5 \times 1.4$ µm$^2$), 3-SNAP (24.0 µA, 50nm, 150nm, $3.5 \times 2.2$ µm$^2$), and 50 nm wide SNSPD (7.2 µA, 50nm, 150nm, $3.5 \times 2.8$ µm$^2$). A mode-locked laser at a wavelength of 1.5 µm with sub-picosecond pulse width and 77 MHz repetition rate was used as the light source in measuring detection efficiency and timing jitter.

## 3. Characterization

As one important motivation of our successive-avalanche architecture, the avalanche current of $n$-*SNAP was expected to be lower than the avalanche current of an $n$-SNAP, therefore, giving more bias margin for paralleling more nanowires to achieve higher signal amplification. As we have discussed, it is difficult to fabricate a well functional $n$-SNAP with $n > 4$, but we were nevertheless able to compare the architectures by using idealized devices in an electrothermal simulation. We ran the simulation several times, varying the bias current in the detection nanowire ($I_{BN}$). Fig. 2(b) shows the output pulses from simulations of an 8-*SNAP with $I_{BN}$ set to $0.70I_{CN}$, $0.75I_{CN}$, and $0.80I_{CN}$. The pulses had three levels, which corresponded to the cases where $A_1$, $A_2$ or $A_3$ was the final avalanche. When $I_{BN}$ was at or below $0.65I_{CN}$, even the pair of $D_1$ and $D_2$ was not triggered. By sweeping the current from $0.75I_{CN}$ to $0.80I_{CN}$, we determined the lowest $I_{BN}$ which was able to trigger $A_3$. This lowest $I_{BN}$, multiplied by a factor of eight, was the minimum avalanche current for the 8-*SNAP and had a value of $I_{AV\_S} = 0.78I_{C8}$, where $I_{C8} = 8I_{CN}$ was the total critical current of the eight nanowires. For simulating the avalanche current of the 8-SNAP ($I_{AV\_C}$) with the same critical current, the choke inductor was set to the same inductance as $L_{S3}$ in Fig.1(c). Applying the same simulation method, we found $I_{AV\_C} = 0.91\ I_{C8}$. Comparing these simulation results, the normalized avalanche current of 8-*SNAP was 13% lower than the normalized avalanche current of 8-SNAP, showing that the successive-avalanche facilitated the avalanche to be happened at low bias current.

To verify the fabricated 8-*SNAP operating in a full avalanche regime, we experimentally measured the avalanche current ($I_{AV\_E}$). We found this value by measuring the device detection efficiency (*DDE*) versus the bias current normalized to its critical current at different incident photon fluxes. At the inflection point of the *DDE* curves, we know that the device is operating in the single-photon regime, and thus we have reached the avalanche current—a full avalanche is occurring from a single detection nanowire firing. As shown in Fig. 2(c), this value was measured to be $I_{AV\_E} = 0.78\ I_{C8}$, in agreement with $I_{AV\_S}$ from our electrothermal simulation. Additionally, we corroborated this measurement by comparing the *DDE* of the 8-*SNAP to a standard, single-nanowire, 50 nm wide SNSPD. Above the normalized bias current of 0.78, the *DDE* curves of the 8-*SNAP overlapped with the curve of the SNSPD as expected from single photon detections in identical nanowires.

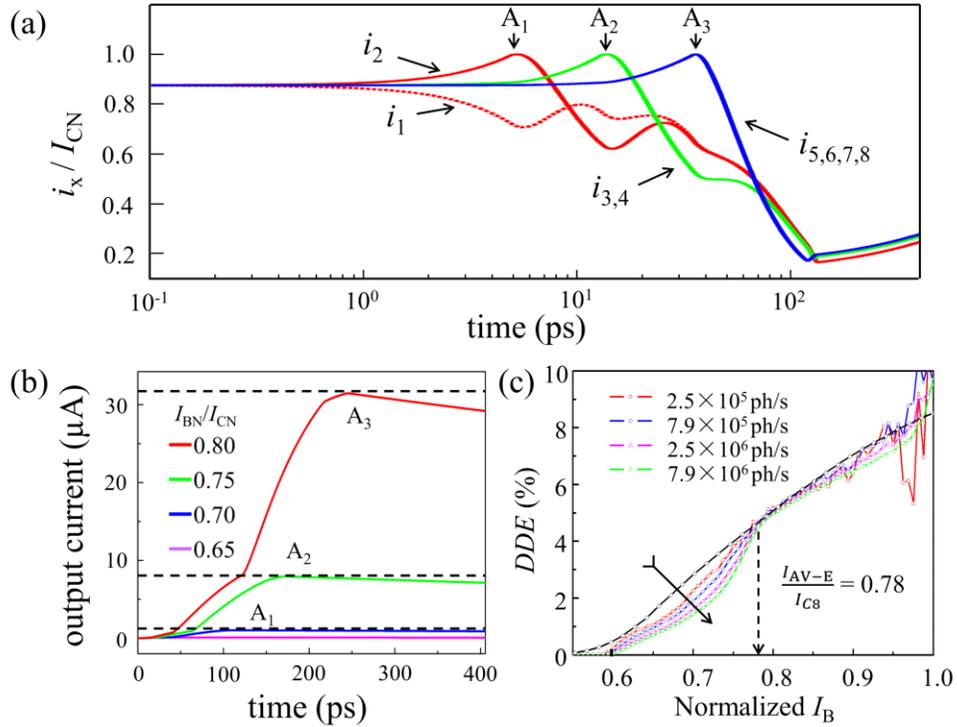

Fig. 2 (a) Electrothermal simulation of the normalized currents through the eight nanowires in an 8-*SNAP. When the current passing through the nanowire exceeds $I_{CN}$, the avalanche processes are triggered, denoted as $A_1$, $A_2$ and $A_3$. (b) Electrothermal simulation of the output pulses at $I_{BN}$ of $0.80I_{CN}$ (red), $0.75I_{CN}$ (green), $0.70I_{CN}$ (blue) and $0.65I_{CN}$ (purple). The dashed lines show the levels of the output current corresponding to the cases that the last avalanche is $A_1$, $A_2$ and $A_3$. (c) *DDE* vs normalized $I_B$ for the SNSPD (black) and the 8-*SNAP illuminated at different incident photon fluxes as shown in the legend. As the photon flux increases depicted by the arrow, the *DDE* curves of the 8-*SNAP shift away from the *DDE* curve of the SNSPD. The dashed arrow shows the inflection point of the *DDE* curves, which is $0.78I_{C8}$ and is the experimental avalanche current of the 8-*SNAP, $I_{AV-E}$. Curves measured under low incident photon fluxes are noisy when $I_B$ is close to the critical current, because the photon detection count rate is lower than the dark count rate, resulting in high background noise.

Ideally, the avalanche current of an 8-*SNAP would be approximately equal to a 2-SNAP. This equivalence should exist because at the first avalanche stage, such as at $A_1$, the current expelled from a detection nanowire only needs to over-bias one neighbor, similar to the two-parallel-nanowire setup in a 2-SNAP. We measured the 2-SNAPs and 3-SNAPs, finding their experimental avalanche currents of $I_{AV-2E} = 0.72\ I_{C2}$ and $I_{AV-3E} = 0.79\ I_{C3}$, where $I_{C2}$ and $I_{C3}$ are the critical currents for the 2-SNAPs and 3-SNAPs, respectively. According to these values, the normalized avalanche current $I_{AV\_E}/I_{C8}$ of the 8-*SNAP was higher than $I_{AV\_2E}/I_{C2}$ of the 2-SNAP and closer to $I_{AV\_3E}/I_{C3}$ of the 3-SNAP, which is slightly different to an ideal equivalence because of the leakage currents in each avalanche stage.

To demonstrate the eight-fold signal amplification of the 8-*SNAP versus a standard SNSPD with the same width of 50 nm, we measured single-shot pulse traces of the 8-*SNAP and the SNSPD. The results are shown in Fig. 3(a). The critical current of the 50 nm wide SNSPD was 7.2 µA and the critical current of the 8-*SNAP was 65 µA. We biased both devices at 83% of their critical currents. The peak values of the 8-*SNAP and the SNSPD pulse were 109.1 mV and 13.5 mV, respectively. This corresponded to an eight-fold enhancement of the signal, proportional to $n = 8$ as expected. This proportionality also shows that the pulse of the 8-*SNAP was the result of a full avalanche sequence containing all the current from all of the detection nanowires, and the successive avalanches worked. Considering the ~100 ns long duration of the 8-*SNAP's output pulse, the system readout noise (peak-to-peak value) was extracted in a 100 ns long time window from the acquired traces before the pulses arrive. The noise was 7.9 mV for the 8-*SNAP and it was 7.3 mV for the SNSPD, which were almost identical and were dominated by the amplifier noise. Therefore, the *SNR* of the 8-*SNAP (109.1/7.9 = 13.8) was also increased by a factor of 8 over the *SNR* of the SNSPD (13.5/7.3 = 1.8). The large *SNR* pulses were easily discriminated from the background noise even with a highly lossy RF cable for connecting. Therefore, the 8-*SNAP could be used as a basic element in an array to increase the amount of pixels with more connections.

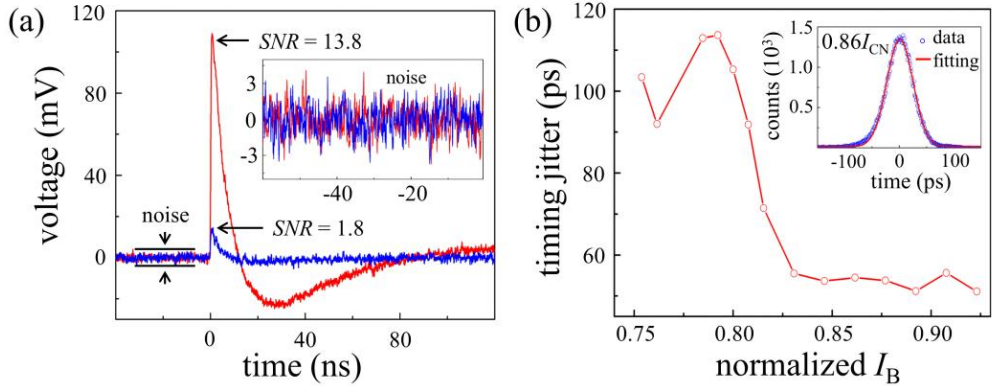

Fig. 3. (a) Traces of a single pulse from the 8-*SNAP (red) and the SNSPD (blue). The signal is extracted from the peak value of the average pulses (not shown here) and the peak-to-peak noise in a time window of 100 ns is taken from the traces before the pulses arrive. The inset figure shows the noise floor of our 8-*SNAP which is nearly identical to the noise floor of the SNSPD. (b) The timing jitter versus normalized $I_B$. The inset figure shows the jitter performance at $0.86I_{C8}$. The delay distribution is fitted to a Gaussian profile.

As shown by the pulse traces in Fig. 3(a), the falling edge of the 8-*SNAP output signal showed an overshoot as the pulse returns to zero. This overshoot indicates a large kinetic inductance of the device which is mainly due to the choke inductors. Reducing the choke inductance could reduce the reset time of the 8-*SNAP, but it also prevented the avalanche formation because the current from the triggered nanowire was more easily to leak to the load instead of injecting into the neighboring nanowires, resulting in an increase of the avalanche current. [12] In the measured circuit, the recovery time constants are set by $L_T/Z_0$, where $L_T \approx L_{S1}$ is the total kinetic inductance of the device and $Z_0 = 50\ \Omega$ is the load resistance that the output pulses apply on. Adding series resistor $R_S$ close to the device, can reduce the time constant to $L_T/(R_S+Z_0)$ and thus can speed up the current recovery in the 8-*SNAP [14]. In an optimized circuit, the 8-*SNAP speed should only be limited by the hotspot cooldown time just like an SNSPD, with a small speed reduction accounted for the delay caused by avalanche latency. There was no afterpulsing in the 8-*SNAP as long as it was not biased very close to the critical current. The bias condition with no afterpulsing existed because the avalanche current of 8-*SNAP was low, showing another advantage of our successive-avalanche architecture.

Since the dynamic processes of the nanowires in an $n$-SNAP are more complex than a SNSPD with single nanowire, we measured the timing jitter of the 8-*SNAP to show that the successive-avalanche architecture does not add significant timing jitter. The timing jitter was extracted by analyzing the variance in time-delay between the synchronized pulses from a fast photodiode and the output pulses from the detectors. [16] The jitter was measured for several values of $I_B$, and the results are shown in Fig. 3(b). Similar to the $n$-SNAPs in conventional architecture, the histogram of the jitter appeared to be asymmetric and showed multiple peaks below the avalanche current. [16] However, when the device was biased above the avalanche current, the histogram of the jitter has a roughly Gaussian profile as shown in the Fig. 3(c) inset. The value of the timing jitter was extracted from the full-width-at-half-maximum (FWHM) of the histogram. For the 8-*SNAP, the jitter had a minimum plateau of ~54 ps, showing negligible dependence on the bias current above 0.82 $I_{C8}$ in the absence of additional timing jitter from voltage noise. [4,5] The jitter was 46 ps for the 2-SNAP biased at 13 μA and it was 53 ps for the 3-SNAP biased at 22 μA. Therefore, the timing jitter of the 8-*SNAP was similar to a 2-SNAP and a 3-SNAP, showing the successive-avalanche architecture didn't give additional timing jitter compared to the conventional design.

**4. Conclusion**

In conclusion, we have demonstrated a parallel SNSPD architecture with successive avalanches. The architecture consisted of detection nanowires and choke inductors arranged in a binary-tree layout. For the demonstrated 8-*SNAP, the output signal and the *SNR* of the output pulse were measured to be eight times higher than the equivalent SNSPD of the same nanowire width. The normalized avalanche current was measured to be 0.78, which was 13% lower than the ideal normalized avalanche current of an equivalent 8-SNAP using conventional design. The detector had a minimum timing jitter of 54 ps, which was independent of the bias current above 0.82 $I_{C8}$, showing that the successive-avalanche architecture didn't deteriorate the timing performance. The binary-tree design can be easily iterated with more nanowires in parallel for greater amplification. The high SNR, low timing jitter, and monolithic integration suggest that this architecture is promising for applications that require larger pulse amplitudes, such as single-photon arrays, photon-number-resolving detectors, or feed-forward on-chip logic for integrated photonic quantum-information processing.

**Acknowledgement:**

The authors would like to thank James Daley and Mark Mondol for technical support. This material is based upon work supported as part of the Center for Excitonics, an Energy Frontier Research Center funded by the U.S. Department of Energy, Office of Science, Office of Basic Energy Sciences under Award Number DE-SC0001088. Qingyuan Zhao would like to thank his financial support (NO.2011619021) from the Chinese Scholarship Council when he was a visiting student in MIT, and the program B (NO.201301B006) for Outstanding PhD candidate of Nanjing University for support during the final stages of manuscript preparation. Adam McCaughan acknowledges support from the iQuISE NSF IGERT fellowship.

**References and links:**
1. E. Knill, R. Laflamme, and G. J. Milburn, "A scheme for efficient quantum computation with linear optics.," Nature 409, 46–52 (2001).
2. N. Gisin and R. Thew, "Quantum communication," Nat. Photonics 1, 165–171 (2007).
3. R. H. Hadfield, "Single-photon detectors for optical quantum information applications," Nat. Photonics 3, 696–705 (2009).
4. Q. Zhao, L. Zhang, T. Jia, L. Kang, W. Xu, C. Cao, J. Chen, and P. Wu, "Intrinsic timing jitter of superconducting nanowire single-photon detectors," Appl. Phys. B 104, 673–678 (2011).
5. L. You, X. Yang, Y. He, W. Zhang, D. Liu, W. Zhang, L. Zhang, L. Zhang, X. Liu, S. Chen, Z. Wang, and X. Xie, "Jitter analysis of a superconducting nanowire single photon detector," AIP Adv. 3, 072135 (2013).
6. A. Divochiy, F. Marsili, D. Bitauld, A. Gaggero, R. Leoni, F. Mattioli, A. Korneev, V. Seleznev, N. Kaurova, O. Minaeva, G. Gol'tsman, K. G. Lagoudakis, M. Benkhaoul, F. Lévy, A. Fiore, G. G. O. L. Tsman, and F. Le, "Superconducting nanowire photon-number-resolving detector at telecommunication wavelengths," Nat. Photonics 2, 302–306 (2008).
7. S. Miki, T. Yamashita, Z. Wang, and H. Terai, "A 64-pixel NbTiN superconducting nanowire single-photon detector array for spatially resolved photon detection," Opt. Express 22, 7811 (2014).
8. Z. Zhou, S. Jahanmirinejad, F. Mattioli, D. Sahin, G. Frucci, A. Gaggero, R. Leoni, and A. Fiore, "Superconducting series nanowire detector counting up to twelve photons," 214, 210–214 (2013).
9. Q. Zhao, A. McCaughan, F. Bellei, F. Najafi, A. De Fazio, A. Dane, Y. Ivry, and K. K. Berggren, "Superconducting-nanowire single-photon-detector linear array," Appl. Phys. Lett. 103, 142602 (2013).
10. M. Ejrnaes, R. Cristiano, O. Quaranta, S. Pagano, A. Gaggero, F. Mattioli, R. Leoni, B. Voronov, and G. Gol'tsman, "A cascade switching superconducting single photon detector," Appl. Phys. Lett. 91, 262509 (2007).
11. F. Marsili, F. Najafi, E. A. Dauler, F. Bellei, X. Hu, M. C. Csete, R. J. Molnar, and K. K. Berggren, "Single-photon detectors based on ultranarrow superconducting nanowires.," Nano Lett. 11, 2048–53 (2011).
12. F. Marsili, F. Najafi, E. Dauler, R. J. Molnar, and K. K. Berggren, "Afterpulsing and instability in superconducting nanowire avalanche photodetectors," Appl. Phys. Lett. 100, 112601 (2012).
13. R. Cristiano, M. Ejrnaes, A. Casaburi, S. Pagano, F. Mattioli, A. Gaggero, and R. Leoni, "Superconducting single photon detectors based on multiple cascade switches of parallel NbN nanowires," in Proc. SPIE 8072, Photon Counting Applications, Quantum Optics, and Quantum Information Transfer and Processing III,, I. Prochazka and J. Fiurášek, eds. (2011), Vol. 8072, pp. 807205–807205–6.
14. J. K. W. Yang, A. J. Kerman, E. A. Dauler, V. Anant, K. M. Rosfjord, and K. K. Berggren, "Modeling the Electrical and Thermal Response of Superconducting Nanowire Single-Photon Detectors," IEEE Trans. Appl. Supercond. 17, 581–585 (2007).
15. A. J. Kerman, J. K. W. Yang, R. J. Molnar, E. A. Dauler, and K. K. Berggren, "Electrothermal feedback in superconducting nanowire single-photon detectors," Phys. Rev. B 79, 100509 (2009).
16. F. Najafi, F. Marsili, E. A. Dauler, R. J. Molnar, and K. K. Berggren, "Timing performance of 30-nm-wide superconducting nanowire avalanche photodetectors," Appl. Phys. Lett. 100, 152602 (2012).